\documentclass[12pt]{article}
\usepackage[dvips]{color}
\usepackage{graphicx}
\begin{document}
\begin{center}
{\LARGE \bf A quantum computer based on \\electrons floating on liquid helium}

\hfill

{\large M.I. Dykman $^1$ and P.M. Platzman$^2$}

\hfill

{\small
$^1$Department of Physics \& Astronomy, Michigan State University,\\
East Lansing, MI 48824, USA\\
$^2$Bell Laboratories, Lucent Technologies, Murray Hill, NJ 07974, USA
}

\end{center}




\abstract{\footnotesize
Electrons on a helium surface form a quasi
two-dimensional system which displays the highest mobility reached in
condensed matter physics. We propose to use this system as a set of
interacting quantum bits. We will briefly describe the system and
discuss how the qubits can be addressed and manipulated, including
interqubit excitation transfer. The working frequency of the proposed
quantum computer is $\sim 1$~GHz. The
relaxation rate can be at least 5 orders of magnitude smaller, for low
temperatures.
}

\section{Introduction}

Finding physical systems that are suitable for quantum computation is
a major challenge, since these systems must consist of a large number
of quantum objects whose states and interactions can be conveniently
manipulated.  In addition, the qubits must be sufficiently isolated
from other degrees of freedom so that they do not disturb the time
dependence of the computational wave function.  It should also be
possible to read out the final state of the system (the result of the
calculation)\cite{Divincenzo-95}.  Ions and neutral atoms in traps
\cite{Zoller-95}, cavity quantum electrodynamic systems
\cite{Turchette-95}, bulk NMR systems\cite{Gershenfeld-97}, quantum
dots\cite{Loss-98}, nuclear spins of atoms in doped silicon devices
\cite{Kane-98}, localized electron spins in semiconductor
heterostructures\cite{Vrijen-00}, and Josephson-junction based
systems\cite{Averin-98} have been proposed, and in a number of cases
proof of the principle has been convincingly demonstrated. Many of
these suggestions and some very recent ideas are discussed in these
Proceedings. However, for all proposed systems there are high
technological and scientific barriers, which must be overcome to
achieve a useful quantum computer.

We believe, based on realistic estimates\cite{Platzman-99}, that there
already exists a system that is a good candidate for a scalable analog
quantum computer (AQC) with an easily manipulated set of qubits.  This
is a system of electrons floating on the surface of superfluid helium.
It is incredibly clean, because the electrons essentially reside in
vacuum, and the closest condensed matter is the superfluid helium,
which is itself very pure, with a microscopically smooth surface.
Electrons on helium display the highest mobility achieved in a
condensed matter system so far\cite{Shirahama-95}.  They have been
extensively studied theoretically and experimentally\cite{Andrei-97},
and are by now well understood.  Our estimates indicate
\cite{Platzman-99} that, under realistically obtainable geometry,
temperature, and magnetic field, the electrons can function as qubits
provided they are confined by micron-size electrodes located below the
helium surface.

The system of electrons on helium has several important advantageous
features. First of all, different states of a qubit have different
electric dipole moments. This makes it relatively simple to address
the qubits and to read out their states. A realistic calculation shows
that the coherence time of the electron states is $\sim 10^5$ times
the reciprocal working frequency of the proposed AQC, which is
determined by the Rabi frequency and the energy of qubit-qubit
interaction. The typical interqubit distance is $\sim 1\,\mu$m, and
therefore fabricating a multi-qubit system should not require
technological breakthroughs.

In what follows we briefly review the physics of electrons on helium,
describe the proposed design of the quantum computer
\cite{Platzman-99}, and discuss how operations on qubits can be
performed.

\section{Electrons on helium}

The geometry of experiments on electrons trapped above the
liquid-helium surface is shown schematically in Fig. 1.  This very
clean system has been recognized as a perfect tool for investigating
many basic concepts of condensed matter physics.  Not only the
electron energy spectrum and scattering mechanisms are well
established\cite{Andrei-97}, but this is one of only a few
well-understood strongly correlated
electron systems. Correlation effects are strong because, although the
system is nondegenerate, the ratio of the Coulomb energy of the
electron-electron interaction to the thermal energy $\Gamma = e^2(\pi
n_e)^{1/2}/k_BT$ usually exceeds 20 ($n_e$ is the electron density;
typically $n_e\sim 10^8$~cm$^{-2}$).  For $\Gamma > 130$, the
electrons form a Wigner crystal\cite{Wigner_expr,Wigner_theory},
whereas for higher temperatures they form a nondegenerate correlated
electron liquid, with unusual transport characteristics
\cite{Dykman-93}.

\begin{center}
\includegraphics[width=12cm]{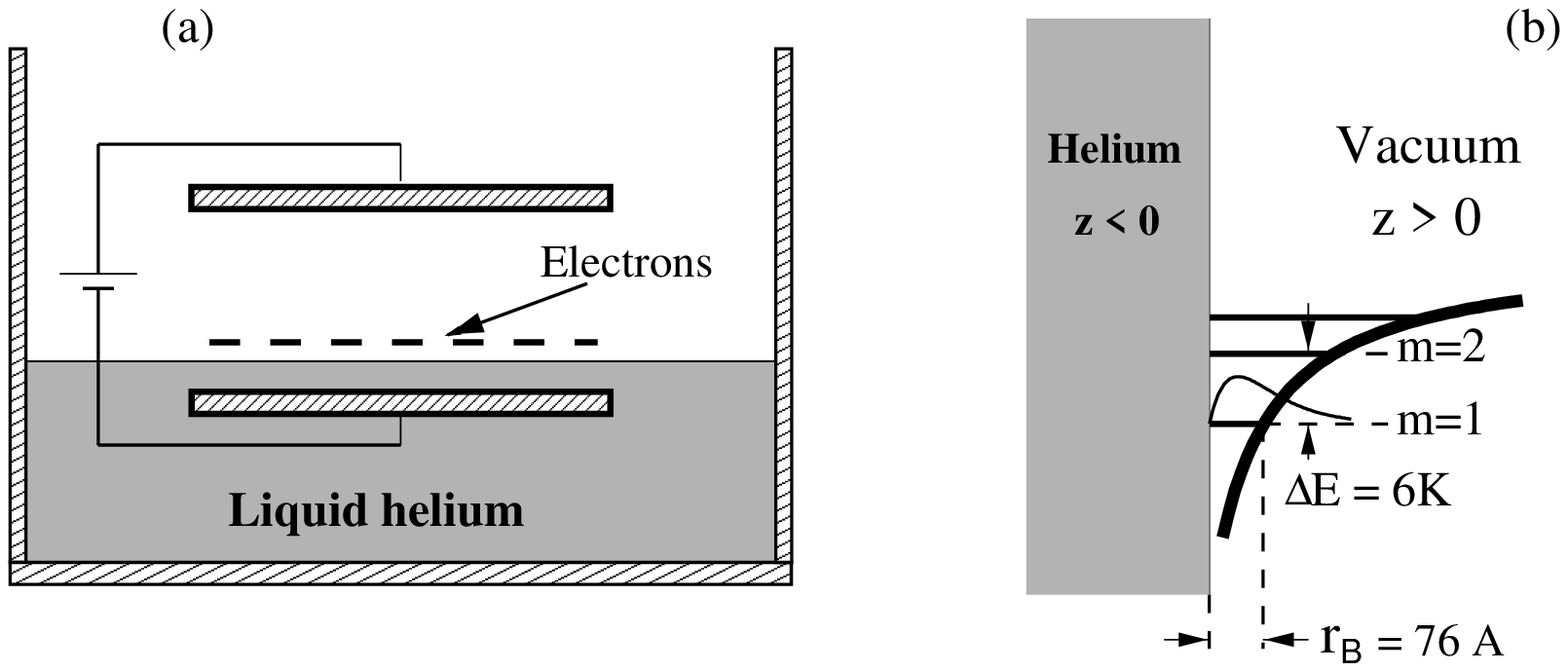}
\end{center}

{\footnotesize Figure 1. (a) Conventional geometry of experiments on
electrons trapped at the helium-vacuum interface.  The electrons are
attracted to the liquid helium by the image potential and an
electrostatic potential of the capacitor, and are free to move in the
plane. (b) The hydrogenic energy levels for electron motion in the $z$
direction transverse to the helium surface, along with a typical $m =
1$ ground state wave function are displayed in the image potential for
$z > 0$.}

\hfill

An electron on a bulk helium film is bound in the direction $z$ normal
to the helium surface by an image potential of the form $V(z) =
-\Lambda e^2/z$ (see Fig. 1),with $\Lambda = (\epsilon -
1)/4(\epsilon + 1)$ ($\epsilon = 1.057$ is the dielectric
constant of liquid helium).  Because there is a barrier of $\approx 1$ eV
for penetration into the helium, the electron $z$-motion is described
by a 1D hydrogenic spectrum. The $m$th state has an energy
$E_m = -R/m^2$, with an effective Rydberg energy $R =
\Lambda^2e^4m_e/2\hbar^2 \approx 8$~K, and effective Bohr radius $r_B =
\hbar^2/\Lambda m_e e^2 \approx 76$~\AA (Fig. 1).

Electronic transitions between the states in Fig.~1b were observed
using microwave spectroscopy\cite{Grimes-76}. The level spacing could
be controlled by a static electric field ${\cal E}_{\perp}$ normal to
the helium surface from the capacitor in Fig.~1a.  This field
Stark-shifted the transition frequencies. The linear Stark shift is
$\approx 1$~GHz/(V/cm), which will be important for
our AQC. The calculated shift was in excellent agreement with the
experiment.  For $T=0.4$~K the
natural linewidth for the $1\rightarrow 2$ phototransition was
estimated from measurements of inhomogeneously-broadened spectra to
be below 15 MHz\cite{Volodin-81}.

For $T < 0.7$~K, the only substantial electron coupling to the outside
world leading to decoherence effects is to thermally excited height
variations $\delta({\bf r},t)$ of the helium surface, where ${\bf r}$ is
the electron in-plane coordinate.  These variations are described as
propagating capillary waves, ripplons.  Ripplons are very soft
excitations: for the wave vectors $k < 5\times 10^5$~ cm$^{-1}$, which
are of interest for the AQC, the ripplon energies are
$\hbar\omega_{\rm r}({\bf k}) < 4\times 10^{-3}$~K.  Therefore, at
$10^{-2}$~K many such ripplons are present, and the corresponding
average mean-square displacement of the surface is determined by
thermal fluctuations, with $\delta_T \approx 2\times 10^{-9}$~cm.
 	
The electron-ripplon coupling energy is 
\begin{equation}
\label{interaction}
H_{\rm e-r}=e\hat{\cal
E}_{\perp}\delta({\bf r},t),
\end{equation}
where $\hat{\cal E}_{\perp}$ is the
effective $z$-directed electric field on the electron.  It is given by
the sum of the externally applied field ${\cal E}_{\perp}$ and the
[nonlocal in ${\bf r}$] field from the image potential $\sim 10^2 - 10^3$~V/cm.

In the lowest hydrogenic level it is the weak coupling to ripplons
which limits the in-plane mobility of the surface state electrons.  It
also determines the dephasing rate of the $m\rightarrow m'$
photoinduced transitions for unconfined electrons.  This dephasing
rate is of the order of the intraband momentum relaxation rate
$\tau^{-1}$.  In the single-electron approximation, $\tau^{-1}=
e^2{\cal E}_{\rm eff}^2/4\sigma$, where ${\cal E}_{\rm eff}$ is the
known characteristic value of the field pressing the electron to the
surface (it is linear in ${\cal E}_{\perp}$), and $\sigma$ is the
surface tension of helium.  This expression is in good agreement with
the measurements \cite{Dahm-Vinen}.  The relaxation rate is record
low: from the recent mobility measurements\cite{Shirahama-95},
$\tau^{-1}\approx 10^7$~s$^{-1}$, in agreement with the upper bound on
the photoabsortpion linewidth cited before.

Since all measurements refer to a strongly correlated electron system,
one may ask if the agreement of the single-electron theory and
experiment is not fortuitous.  However, in a broad range of $T$ and
$n_e$, the single-electron transport theory is essentially applicable
in the absence of a magnetic field \cite{Dykman-93}.  This is no
longer true in the presence of a field $B_{\perp}$ normal to the
electron layer, which is also of interest for the AQC.  It turns out
that transport in the correlated electron system can be analyzed in a
non-perturbative way\cite{Dykman-93}.  The many-electron
conductivity displays a specific dependence on the parameters of the
system, and in particular, it is a non-monotonic function of
$B_{\perp}$. The theory is in full agreement with the
experiment\cite{Lea-97}, which reassures one that our understanding
of the electron system on helium is adequate.

Besides intraband relaxation, the coupling to ripplons gives rise to
transitions from the excited $m = 2$ to the ground $m = 1$ hydrogenic
band in Fig. 1.  A simple estimate for the interaction
(\ref{interaction}) gives the relaxation rate $T_1^{-1}\sim
(R/\hbar)(\delta_T/r_B)^2$.  Since $\delta_T/r_B \sim 10^{-3}$, the
rate $T_1^{-1}$ is $\sim 10^{-6}$ of the transition frequency $\sim 120$ GHz.
This suggests using the lowest two hydrogenic levels of an individual
electron as a convenient qubit, whose state can be changed by the
application of a resonant microwave field.  

For a field amplitude $E_{\rm RF} = 1$ V/cm, the Rabi frequency
$\Omega =\left|eE_{\rm RF}\langle1|z|2\rangle\right|/\hbar \sim
eE_{\rm RF}r_B/\hbar$ of the oscillations of a resonantly driven
electron between the lowest and the first excited state is about
$10^9$ s$^{-1}$.  This is the working frequency of the quantum qubit
we are suggesting.  Even a laterally unconfined electron has $\Omega
T_1 > 10^4$.

\section{Qubits using electrons on helium}

Recently we suggested\cite{Platzman-99} that electrons on helium can
be used as addressable qubits if they are confined laterally.  The
geometry is shown schematically in Fig.~2.  The lower plate of the
capacitor that holds electrons at the helium surface is patterned and
forms a set of individually addressable micro electrodes.  To make the
frequency of resonant energy exchange between neighboring electrons of
the order of $\Omega\sim 10^9$~s$^{-1}$, the distance $d$ between the
dots should be $\sim 0.5\,\mu$m.  Consequently, the electrodes should
be submerged at $h\sim d\sim 0.5\,\mu$m beneath the helium surface.
This geometry allows accommodation of $\geq 10^8$ qubits, for a
typical area of the helium surface 1~cm$^2$.

\hfill

\parbox{3.0in}{
\begin{center}
\includegraphics[width=8.0cm]{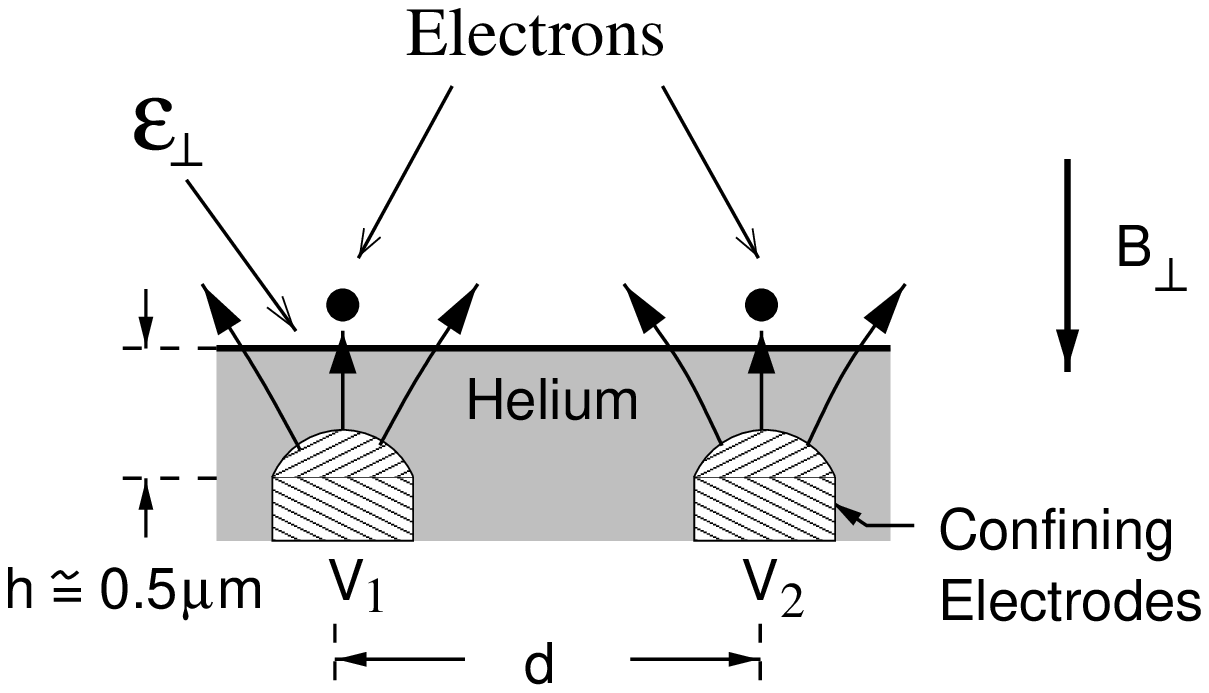}
\end{center}

}\ \hfill
\parbox{2.0in}{\footnotesize Figure 2. Electrons on helium above a patterned
substrate. The rough dimensions ($h\approx d$), the shape of the field
lines, and the gate voltages are included. The average distances of
the electron from the helium surface in the hydrogenic states
$|1\rangle$ and $|2\rangle$ are $\approx 11$~nm and $\approx 45$~nm,
respectively, for small pressing field ${\cal E}_{\perp}$.}
\addtocounter{figure}{1}

\hfill

The electrode potentials $V_n$ create both out-of-plane and in-plane
fields on the electrons. They can be easily found for a spherical
electrode of radius $R$. If the center of the sphere is at the
distance $h$ below the surface, the $V_n$-dependent part of the
pressing field ${\cal E}_{\perp}$ is ${\cal E}_n\approx V_nR/h^2$. Then, for
the geometry in Fig.~2, the change of the Bohr frequency of the
$1\rightarrow 2$ transition by 1 GHz should occur for the increment 
$\Delta V_n \sim 0.1-1$~mV [with account taken of the
neutralizing background].  This allows one to make a {\bf one-qubit
gate} by tuning the Bohr frequency of the targeted electron to the
frequency of the externally applied microwave radiation.  Depending on
the duration of the radiation pulse $T_{\rm RF}$ with respect to the
Rabi frequency $\Omega$, the qubit will be put into an arbitrary
superposition of the ground and excited states.  In fact, an optimal
arrangement is to combine a radiation pulse with the pulse of the
control voltage $V_n$, which tunes the qubit into resonance for a
desired amount of time.

Electron confinement gives rise to a further dramatic increase of the
lifetime of the excited state $T_1$, and makes the coherence time
$T_2$ very large.  The spectrum of a single pinned electron consists
of a ladder of discrete in-plane energy levels connected with the
ground and excited hydrogenic states in Fig.~1. The level spacing
$\hbar\Omega_{\parallel}$ can be estimated for spherical electrodes,
in which case $m\Omega_{\parallel}^2 \approx e^2R(h^2-R^2)^{-2} +
eRV_nh^{-3}$. This spacing is not commensurate with $E_2 - E_1$.
Therefore a transition from the state $m = 2$ requires emission of a
ripplon with energy $\sim \hbar\Omega_{\parallel}$, which is much
higher ($\sim 0.4$~K) than $\hbar\omega_{\rm r}$.  
We can further suppress one-ripplon decay processes if we apply a
magnetic field $B_{\perp}$ normal to the electron layer. It will
open up gaps $\hbar e B_{\perp}/mc$ in the in-plane excitation
spectrum, which are $\approx 2$~K for $B_{\perp} = 1.5$~T.

The dephasing rate $T_2^{-1}$ of a confined electron is determined by
a kind of quasi-elastic scattering of thermally excited ripplons,
which is similar to the single-electron dephasing rate in quantizing
magnetic fields\cite{Dykman-78}. It contains a factor
$(\delta_T/r_B)^4\sim 10^{-11}$ and is estimated\cite{Platzman-99} to
be $T_2^{-1}< 10^{4}$~s$^{-1}$ for $T=10$~mK. The ratio of the working
frequency to the relaxation rate is $\Omega T_2 > 10^5$.  Other
sources of noise and fluctuations can be reduced to the level where
they give an even smaller dephasing rate\cite{Platzman-99}.

{\bf Two-qubit gates} are often discussed in terms of pairing qubits
in turn, in isolatation from the rest of the system, and manipulating
them in a few fundamental ways. However, in real physical systems
interactions are not easily turned on and off.  Therefore we would
argue that the manipulation of interactions in the whole interacting
many qubit system, and the final measurement or collapse of the wave
function must be thought of as part of the ultimate computation
process. In our computer the qubit-qubit interactions come from the
charge on the electrons and are always there. The important ones, as
regards the $z$-motion, are dipolar with energy  given by
$\sum\nolimits_{n \neq m} e^2z_nz_m/2d_{nm}^3$,
where $d_{nm}$ is the distance between the $n$th and $m$th electrons.

The interelectron coupling allows us to perform both CNOT and swap
operations. The latter can be done by sweeping the Bohr frequencies
$\Omega_n=[E_2^{(n)}-E_1^{(n)}]/\hbar$ of the electrons past each
other, as a result of the voltage change on the underlying
electrodes. At resonance, an excitation will be transferred between
nearest neighbors over the time $\sim \Omega_{\rm sw}^{-1},
\;\Omega_{\rm sw}= e^2|\langle 1|z|2\rangle|^2/\hbar d^3$. For the
parameters in Fig.~2, this time is $10^{-9}$~s.

\hfill

\noindent
\parbox{3.0in}{
\begin{center}
\includegraphics[width=9cm]{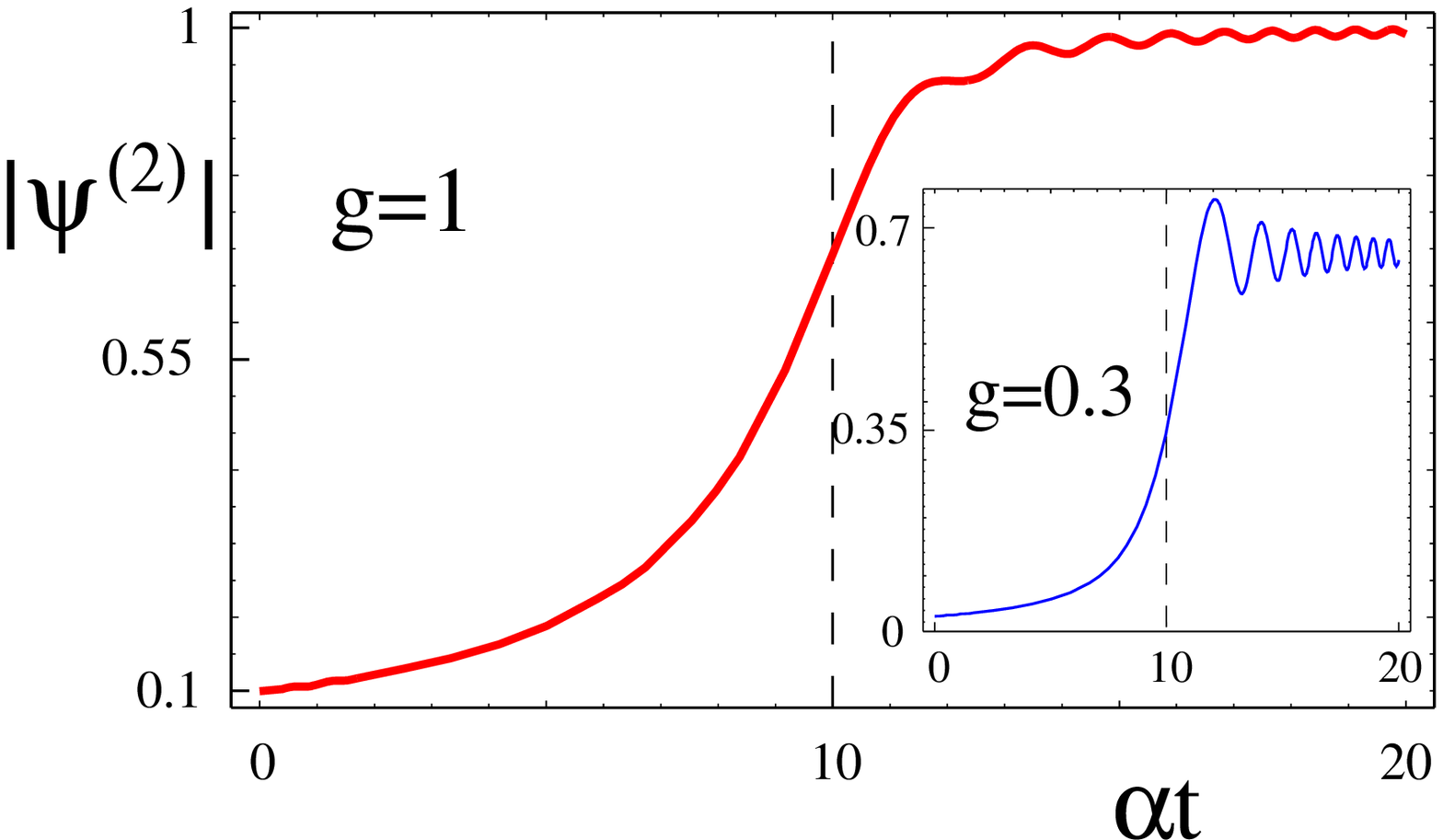}
\end{center}

}\ \hfill
\parbox{2.0in}{\vspace*{-0.1in}\footnotesize Figure 3. Excitation transfer between the qubits
1 and 2, which are initially ($t<0$) in the excited and ground states,
respectively. The Bohr frequencies of the qubits are linearly changed
in time, $\Omega_{1,2}(t)=\Omega_{1,2}(0) \mp\alpha t$ and become
equal to each other for $\alpha t=10$. Solid curves show the amplitude
of the excited state of the qubit 2 for different effective coupling
parameters $g=\Omega_{\rm sw}/\alpha^{1/2}$.}
\addtocounter{figure}{1}

\hfill

Kinetics of excitation transfer between two qubits for linear change
of the electrode potentials $V_{1,2}$ is illustrated in Fig.~3. The
amplitude of the initially occupied excited state of the qubit 1
approaches $\exp(-\pi g^2)$ for large time. Therefore, by adjusting
the rate of the potential change, excitation transfer can be made
exponentially efficient.

A simple way of measuring the state of a qubit is to apply a pulse of
a negative voltage to the underlying electrode. The electron may then
tunnel away from the surface. For the geometry in Fig.~1, the
tunneling was investigated before\cite{Andrei-97,Goodkind-93}. The
tunneling rate is exponentially larger if the electron is in the
excited rather than the ground state. This allows one to detect the
electron state\cite{Platzman-99}. Alternatively, the electron states
may be detected by attaching single electron transistors to the
control electrodes.

The system of two-dimensional electrons on helium is unique in the
context of large AQC systems. It is scalable, easily manipulated, and
has perfectly acceptable decoherence times. It has been carefully
investigated theoretically and experimentally, and there seems to be
no existing technological barriers present for making an AQC using it.

\section*{Acknowledgments}
MID acknowledges support from the NSF through grant No. ITR-0085922.

\end{document}